\documentclass[pra,twocolumn]{revtex4}

\newcommand{\QKD}{{\sc QKD}}
\newcommand{\PNS}{{\sc PNS}}
\usepackage{epsfig}
\begin{document}

\title{Quantum key distribution with realistic states: photon-number
statistics  in the photon-number splitting attack}

\author{Norbert L\"utkenhaus}
\affiliation{Quantum Information Theory Group, Center for Modern Optics (ZEMO), Universit\"at
Erlangen-N\"urnberg, Staudtstr. 7/B2, D-91058 Erlangen,
Germany}
\email{luetkenhaus@kerr.physik.uni-erlangen.de}

\author{Mika Jahma}
\affiliation{Helsinki Institute of Physics, 
  PL 64, FIN-00014 Helsingin yliopisto,
  Finland}

\date{\today}

\begin{abstract}
Quantum key distribution can be performed with practical
signal sources such as weak coherent pulses. One example of such a
scheme is the Bennett-Brassard protocol that
can be implemented via polarization of the signals, or equivalent
signals.  It turns out that the most powerful tool
at the disposition of an eavesdropper is the photon-number splitting
attack. We show that this attack can be extended in the relevant
parameter regime such as to preserve
the Poissonian photon number distribution of the combination of the
signal source and the lossy channel.
\end{abstract}
\maketitle

\section{Introduction}
Quantum Key Distribution (\QKD) allows to generate a long secret shared key
between two parties, conventionally named Alice and Bob, from a short
initial secret key. Part of that newly generated key can then be used
up by sending an unconditionally secure secret message via the one-time
pad, also called Vernam
cipher \cite{vernam26a}. The remaining part is
retained to repeat the \QKD\ protocol to generate new key. The first
complete protocol is that by Bennett and Brassard, BB84  \cite{bennett84a},
although Wiesner formulated basic ideas earlier \cite{wiesner83a}.

In ideal \QKD\ protocols we are required to use particular states for
which the preparation is  beyond our present experimental capability,
such as single
photon states on which we can imprint signals in form of specific
polarizations. For example, for the BB84 protocol we would use two pairs
of orthogonal polarizations, e.g. horizontal/vertical linear polarization and
right/left circular polarization. 

Recently it has been proven that one can use realistic signal sources
such as weak laser pulses polarized in the four signal polarizations
to perform \QKD\ even in the presence of loss and noise in the quantum
channel. Indeed, in most experiments demonstrating the technique
required for \QKD\ this signal source has been used
\cite{chiangga99a,bourennane99a,marand95a,townsend98a,hughes00a,ribordy00a}.
For eavesdropping attacks on such signals, the security of the BB84
protocol in a realistic setting has been explored regarding attacks on
individual signals in \cite{nl00a}.  The proof of unconditionally
security of \QKD\ with the BB84 protocol in this framework has been
presented in \cite{inamorisub01a}. It turns out, that the combination
of multi-photon signals of the source, such as weak laser pulses,
together with loss in the quantum channel in the presence of errors leads
to limitations of rate and distance that can be covered by those
techniques. These restrictions are due not only to the proving
techniques but are of fundamental nature \cite{brassard00a}, at least
in a conservative approach to security where all errors and losses are
assumed to be due to eavesdropping activity.

The limitation comes from the fact that the combination of multi-photon
signals of the source and loss in the transmission line opens the door
for a powerful eavesdropping attack, the Photon Number Splitting
(\PNS) attack that was first mentionend in \cite{huttner95a}. The
basic step is that a signal consisting of two or more photons
(multi-photon signal) can be split via a physical interaction
\cite{nl00a} by an eavesdropper (called Eve) such that Eve retains one
photon and Bob receives the other photons such that the polarization
of both parts remains undisturbed. The photon in Eve's hand will
reveal its signal polarization to Eve if she waits long enough until
she learns the polarization basis during the public discussion part of
the BB84 protocol. In the presence of loss, this attack can put Eve in
a position that she knows the complete information about all signals
received by Bob and no secure key can be generated. This is the case
if the loss is strong enough, such that Bob expects to receive less
signals than the signal source prepares multi-photon signals. Then Eve
can replace the lossy quantum channel by an ideal one, block all
single-photon signals and use only multi-photon signal to match Bob's
expectation of non-vacuum pulses. If the loss is not high enough for
this, then Eve can block only a fraction $b$ of the single-photon
signals, but she can perform some optimal eavesdropping attack on the
remaining single-photon pulses. This constitutes her optimal attack
\cite{nl00a}. Despite this powerful attack, in this situation, if the
error rate is not too high, Alice and Bob can establish a secure key,
as has been shown in \cite{nl00a,inamorisub01a}, in the standard BB84
protocol where no photon-number statistics is monitored.
Note that the \PNS\ attack can be well approximated with linear optics
only \cite{calsamiglia02a}.

One remaining open question is whether Alice and Bob might be able to
detect that Eve performed the \PNS\ attack. After all, the photon
number statistics changes under the \PNS\ attack as described
above. In this paper, we will show that is is possible to extend the
\PNS\ attack such that the photon number statistics, as seen by Bob,
is indistinguishable from that resulting from weak laser pulses and a
lossy channel. This result holds in the relevant paramter regime of
mean photon number $\mu$  of the Poissonian photon number distribution of the
weak laser pulse and of the transmission factor $\eta$ of the quantum
channel. The extension of the PNS attack allows the eavesdropper to
remain undetected even if Alice and Bob measure the photon number
distribution via coincidence rates in
Bob's photodetectors.

\section{Extended \PNS\ attack}
We consider a photon source emitting signals with a Poissonian
photon number distribution with mean value $\mu$. Weak laser pulses
are well described by Fock states with this photon number
distribution, but our analysis can be extended to other distributions. The quantum
channel is described by a single-photon transmission efficiency
$\eta$. Then
we find at Bob's end of the quantum channel again a Poissonian photon
number distribution with mean photon number $\mu \eta$, that is
\begin{equation}
P_{loss}[n] = \frac{(\eta \mu)^n}{n!} \exp[- \eta \mu] \; .
\end{equation}
On the other hand, the \PNS\ attack described above will give another  photon
number statistics. Let Eve perform the photon number splitting attack
in which she blocks a fraction $b$ of the single-photon signals. Then
we find a resulting photon number distribution that is not Poissonian, namely
\begin{equation}
P_{PNS}[n] = \left\{
\begin{array}{cl}
(1+b \mu)\exp[- \mu]  & n=0 \\
\left((1-b)\mu+\mu^2/2\right)\exp[- \mu]  & n=1 \\
\frac{ \mu^{n+1}}{(n+1)!} \exp[- \mu] & n>1 
\end{array} \right.
\end{equation}
To match the number of vacuum signals, we adjust $b$ such that
$P_{loss}[0] = P_{PNS}[0]$. This leads to the expression
\begin{equation}
b_{match } = \frac{1}{\mu} \left( \exp[\mu(1-\eta)]-1 \right)\;.
\end{equation}
We find $b_{match }=0$ for $\eta=1$, while $b_{match }=1$ for $\eta=1-\frac{1}{\mu}
\ln[1+\mu]$. This last point corresponds precisely to the situation
where $1-P_{loss}[0] = 1-(1+\mu)\exp[-\mu]$, that is, the number of
non-vacuum signals arriving at the end of the lossy quantum channel is
equal to the number of multi-photon signals emanating from the
source. For values of $\eta$ between these two extreme values, $b$
takes on values in the interval $(0,1)$, and this is the regime we are
dealing with.

With this choice the photon number distribution after Eve's attack
takes the form
\begin{equation}
P_{match}[n] = \left\{
\begin{array}{cl}
\exp[- \eta \mu]  & n=0 \\
(1+\mu+ \mu^2/2) \exp[-  \mu]- \exp[- \eta \mu] & n=1\\
\frac{ \mu^{n+1}}{(n+1)!} \exp[- \mu]  & n>1 \; .
\end{array} \right.
\end{equation}
This photon number distribution is not Poissonian. The question is
whether Eve can make it Poissonian without loosing any advantage of
the PNS attack. It is easy to come up with a possible solution: Eve
can extract not only one, but two or more photons from pulses
depending on the photon number in each pulse. With this method it is
possible to redistribute probabilities from higher to lower photon
numbers, but not the other way round. Therefore the necessary and
sufficient condition for the redistribution to be possible is that
\begin{equation}
\label{rediscon}
\sum_{i=0}^n P_{loss}[i] \geq \sum_{i=0}^n P_{match}[i]
\end{equation}
is satisfied for all $n$. This condition that the change of
probability from any high-photon number part to a low-photon number
part goes in the right direction.

In order to work as an eavesdropping attack, we need to make sure that
the number of non-vacuum signals remains unchanged. Indeed, in our
extended strategy we will never take 'the last photon' out of a pulse, so the
number of non-vacuum  signals does not change. This guarantees that the
information gain by Eve on the signals remains unchanged. The only
change is that of the photon number statistics of the signals arriving
at Bob's end of the quantum channel.

\section{Evaluation of extended \PNS\ attack}
In this section we will show that the conditions (\ref{rediscon}) is
satisfied in a parameter regime that we will show in the following
section to be  relevant to practical \QKD. Thus
we show that Eve can mimic Poissonian photon number distribution even
while performing the \PNS\ attack via our extension.

We define the $d_n$ as the difference of the two probability
distributions for given $n$, so that we have
\begin{equation}
d_n= P_{match}[n]-P_{loss}[n]\; .
\end{equation}
Note that $d_0=0$ due to the matching via the blocking parameter
$b$. We will show that in a relevant paramter regime we find that $d_n
\leq0$ for $n \in[1,n_l]$ and $d_n \geq 0$ for $n \in[n_l+1,\infty]$. This
is sufficient (though not necessary)  to fulfill the conditions
(\ref{rediscon}). With other words, with increasing value of n, the
difference $d_n$ vanishes for $n=0$, then takes negative values until
for $n\geq n_l+1$ it turns positive. 

Let us first show by induction that once the function turned positive,
it will not turn back negative for $ n \geq 2$ and the parameter
regime $\eta \leq 3/4$. Assume that $d_n \geq 0$ for $n\geq 2$. This
means
\begin{equation}
\frac{ \mu^{n+1}}{(n+1)!}
\exp[- \mu]\geq\frac{(\eta \mu)^n}{n!} \exp[- \eta \mu]  \; .
\end{equation}
Then it follows that
\begin{eqnarray}
\lefteqn{d_{n+1}}\\
 & = & \frac{ \mu^{n+1}}{(n+1)!} \exp[- \mu]
 \frac{\mu}{n+2}-\frac{(\eta \mu)^n}{n!} \exp[- \eta \mu] \frac{\mu
 \eta}{n+1} \nonumber\\
& \geq & \frac{(\eta \mu)^n}{n!} \exp[- \eta \mu] \left(\frac{\mu}{n+2}-\frac{\mu
 \eta}{n+1}\right) \nonumber \\
& \geq & 0  \mbox{\hspace{1cm}for}\; \eta \leq \frac{3}{4} \; . \nonumber
\end{eqnarray}
We do not need to prove directly that there is some $d_{n_l} \geq 0$
with $n_l \geq 2$. Instead, we will show that in a suitable parameter regime $d_1
\leq 0$. That proves together with the normalization
$\sum_{n=0}^\infty d_n =0$ and $d_0=0$ that there must be some
positive $d_n$ for $n \geq 2$. With other words, for $\eta < 3/4$ we
find that $d_1\leq0$ is a sufficient condition to allow a redistribution
of the photon number distribution after the \PNS\ attack to make it
Poissonain without changing the flow of information between the
parties. 

\begin{figure}
\epsfbox{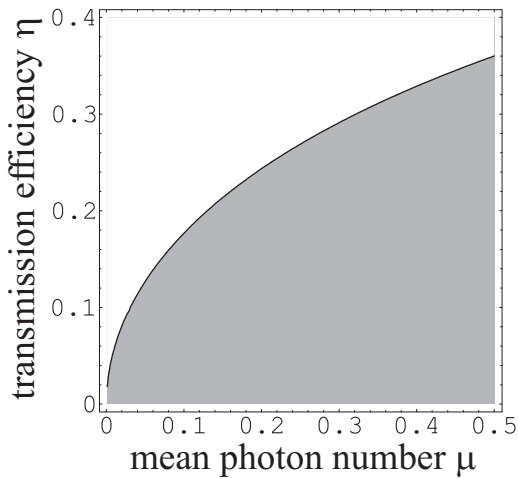}
\caption{\label{d1} As a function of the mean photon number $\mu$ and
the transmission efficiency $\eta$ we see the area (grey shade) where the original
\PNS\ attack yields less single photon signals than the corresponding
lossy channel. In this region the extended \PNS\ attack is successful
in that it mimics the lossy channel not only in the fraction of
non-vacuum signals but also in the whole photon number statistics.} 
\end{figure}

The required condition $d_1 \leq0$ can be analyzed numerically only. It
is given after a regrouping as 
\begin{equation}
(1+\mu+ \mu^2/2) \exp[-  \mu]- (1+ \eta \mu) \exp[- \eta \mu] \leq 0
\; .
\end{equation}
The first term describes the fraction of signals containing $0$ or $1$
photons after the original \PNS\ attack while the second term
describes the target value for this fraction. Note that we fixed the
fraction of vacuum signals so that this is, indeed, a statement about
the fraction of single-photon signals. The region where $d_1$ is
negative is plotted in figure \ref{d1}.

\begin{figure}
\epsfbox{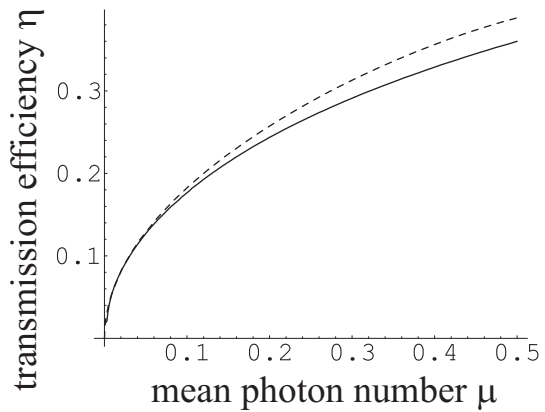}
\caption{\label{approx} We compare the approximation (dashed line) for the critical
value $\eta_0$ for which $d_1$ vanishes with a numerical solution of
$d_1[\eta_0]=0$ (solid line).} 
\end{figure}

The borderline shown in figure \ref{d1} can be evaluated more closely
in the typical regime were $\mu, \eta \ll 1$. If we expand $d_1$ in
$\mu, \eta$ and neglect terms $\eta^k \mu^l$ with $k+l > 4$, then we
obtain
\begin{equation}
\label{d1approx}
d_1 \approx \mu^2/2 \left( -\mu/3+ \mu^2/4 + \eta^2\right)
\end{equation}
so that we find that the value $\eta_0$  for which $d_1$ vanishes can be
approximated by
\begin{equation}
\eta_0 \approx \sqrt{\mu/3-\mu^2/4}
\end{equation}
As we see from Fig.~\ref{approx}, this is a good approximation for low
photon numbers. 
Note that the lowest order of the approximation (\ref{d1approx}) is
given by $d_1 \approx - \mu^3/6$ which guarantees that the extended
\PNS\ attack is always successful as long as $\mu$ and $\eta$ are
small. This is mirrored in the infinitely steep rise of the limiting
line shown in figure \ref{approx}. 

\section{Application to security proofs and discussion}
From the security proofs \cite{inamorisub01a,nl00a} we know the
optimal choice of the mean photon number $\mu$ from the point of view
of Alice and Bob. This optimization can be understood starting from
the idea that Alice and Bob would like to optimize the gain rate of
the \QKD\ process. This gain rate $G$ is bounded in a conservative scenario from above as $G \leq
\frac{1}{2}\left(S_m-p_{exp}\right)$ with $S_m$ as the multi-photon
probability of the source and $p_{exp}$ as the fraction of non-vacuum
signals detected by Bob. The factor $1/2$ comes from a sifting state
of the \QKD\ protocol and is specific for the BB84
protocol. For other polarization based protocols such as the six-state
protocol \cite{bruss98a} we find other factors, however, the reasoning
is independent of this factor. For our Poissonian distributed signals,
we find
\begin{equation}
\frac{1}{2}\left(S_m-p_{exp}\right) = \frac{1}{2}\left(\exp[-\mu \eta]
- (1+\mu) \exp[-\mu]\right)
\end{equation}
and this expression is optimized for small values of $\eta$ by
$\mu_{opt} \approx \eta$. As it turns out, that value remains
approximately optimal in a detailed analysis taking other error
sources into account \cite{inamorisub01a,nl00a}. In typical
experiments we find that even higher photon numbers than the optimal
ones are used. That pushes the working point even further away from
the critical line in figure \ref{d1}.

The statements above are made for a conservative scenario where all
loss and all errors are attributed to Eve. It will be desirable to
extend this analysis by making assumptions stating that Eve cannot
change the dark count rate of Bob's detectors and cannot increase
the detection efficiency. It turns out that this extension is not
trivial at all. The analysis in our paper, however, puts some
bounds on the results of such an analysis. Once Eve can perform an
extended \PNS\ attack such that her action mimics a lossy quantum
channel both in the transmission efficiency and the Poissonian photon
number statistics, and she can block all single-photon signals, then
the transmission cannot be secure. 

At the heart of our paper is the statement that stochastic processes
such as loss in a quantum channel can be explained by a rather
cunning strategy of a third party, for example an eavesdropper. It
means that the eavesdropper can access some preferred signals while
suppressing others such that the resulting action is indistinguishable even
in principle from a normal lossy channel for given input signals. It
is unclear so far what input signals have to be chosen to make such a
situation impossible.

\section{Acknowledgement}
We gladly acknowledge stimulating discussion with Mioslav Du\v{s}ek
and John Calsamiglia. N.L. thanks Marcos Curty and Peter van Loock for
their critical discussion of this article. This work has been
supported under Project NO. 43336 of the Academy of Finland and by the
German Reseach Council (DFG) via the Emmy-Noether Programme.

\end{document}